\documentclass[twoside,symmetric]{tufte-handout}
\usepackage{hyperref}
\usepackage{mathrsfs}
\usepackage{bibentry}
\usepackage{eufrak}
\usepackage{enumerate}
\renewcommand{\today}{\number\day \space%
\ifcase \month \or January\or February\or March\or April\or May%
\or June\or July\or August\or September\or October\or November\or December\fi \space%
\number \year} 
\renewcommand{\doi}[1]{\textsc{doi}: \href{http://dx.doi.org/#1}{\nolinkurl{#1}}}

\newcommand{\B}{\rule[-1.2ex]{0pt}{0pt}} 

\title{Colloquium: A Century of Noether's Theorem}

\author[Chris Quigg]{Chris Quigg\hfill{\small email:quigg@fnal.gov~~~~~~}\\
{\normalsize Theoretical Physics Department\\ Fermi National Accelerator 
Laboratory\\ P.O. Box 500, Batavia, Illinois 60510 USA}}
\date{{\today}\hfill {\texttt{\hbox{\small FERMILAB-PUB-19-059-T~~~~}}}} 

\usepackage{graphicx} 
  \setkeys{Gin}{width=\linewidth,totalheight=\textheight,keepaspectratio}
  \graphicspath{{graphics/}} 
\usepackage{amsmath}  
\usepackage{booktabs,adjustbox} 
\usepackage{units}    
\usepackage{multicol} 
\usepackage{lipsum}   
\usepackage{fancyvrb} 
  \fvset{fontsize=\normalsize}

\begin{document}

\maketitle
\begin{marginfigure}[112pt]
\centering
\includegraphics[width=0.75\columnwidth]{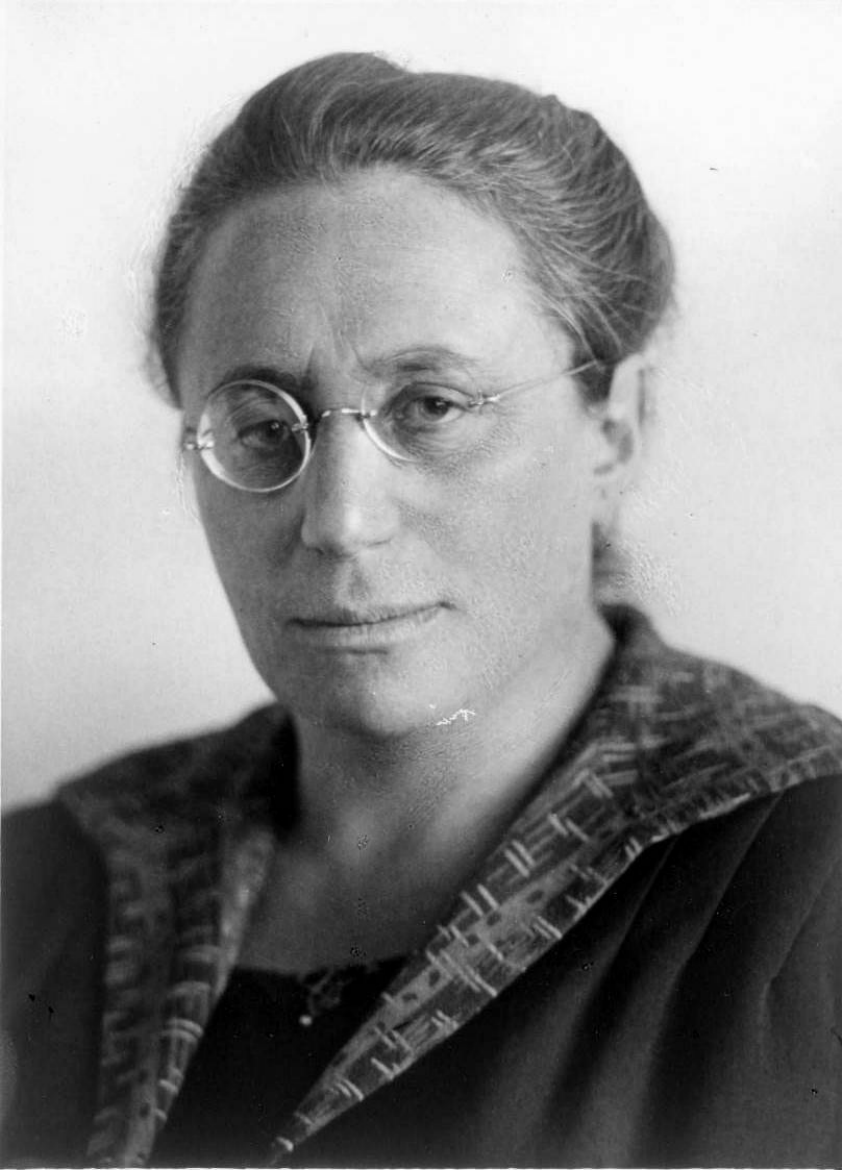}
\break{\href{http://triptych.brynmawr.edu/cdm/singleitem/collection/BMC_photoarc/id/163/rec/1}{Bryn Mawr College Special Collections}}
\end{marginfigure}
\begin{abstract}
\noindent
In the summer of 1918, Emmy Noether published the theorem that now bears her name, establishing a profound two-way connection between symmetries and conservation laws. The influence of this insight is pervasive in physics; it underlies all of our theories of the fundamental interactions and gives meaning to conservation laws that elevates them beyond useful empirical rules. Noether's papers, lectures, and personal interactions with students and colleagues drove the development of abstract algebra, establishing her in the pantheon of twentieth-century mathematicians. This essay traces her path from Erlangen through G\"ottingen to a brief but happy exile at Bryn Mawr College in Pennsylvania, illustrating the importance of ``Noether's Theorem'' for the way we think today. The text draws on a colloquium presented at Fermilab on 15 August 2018.
\end{abstract}


\newthought{On the twenty-sixth of July in 1918,} Felix Klein gave a presentation to the Royal Academy of Sciences in Göttingen$\,$\footnote{Felix Klein is known to popular scientific culture for his conception of the Klein surface (\emph{Fläche})---mistranslated as the Klein bottle (\emph{Flasche}).}. The paper he read had been dedicated to him on the occasion of his Golden Doctorate, the fiftieth anniversary of his Ph.D., by a young colleague named Emmy Noether. The centennial of this paper$\,$\cite{Noether:1918zz,Noether2011}, which contains two theorems that have had an extraordinary impact on physics, including particle physics, for a hundred years, provides the occasion for this commemoration.

It was a busy week in G\"ottingen, and especially for Felix Klein. Not only was he celebrating his \emph{Doktorjubil\"aum}, he had given a paper$\,$\cite{Klein1918} of his own the week before explaining how he and David Hilbert were coming to terms with the idea of energy conservation in Einstein's General Theory of Relativity. They were puzzling over an observation that within the General Theory, what normally is a constraint of energy conservation appears as an identity. How then could it constrain anything? This was the problem on which he had asked Emmy Noether's help. Hilbert is revered among mathematicians for the twenty-three problems$\,$\cite{HilbertProbs} he posed in 1900 and known to physicists for the Courant--Hilbert tomes on methods of mathematical physics.

\begin{marginfigure}
\href{http://www.digizeitschriften.de/dms/img/?PID=GDZPPN00250510X}{\includegraphics{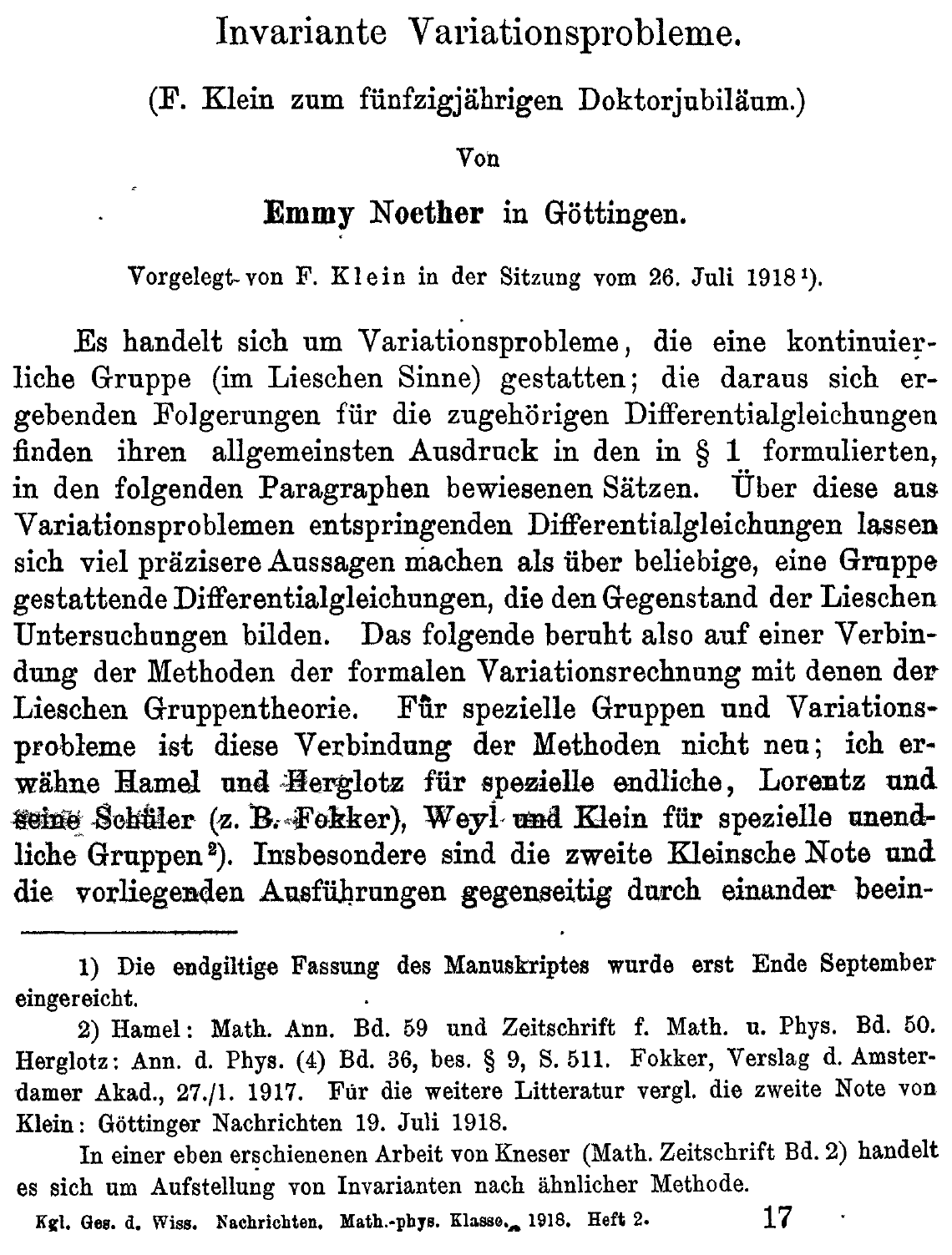}}
\caption{\emph{Invariante Variationsprobleme}\label{fig:ivp}}
\end{marginfigure}

A few days later, on July 23, Emmy Noether summarized the content of her two theorems before the German Mathematical Society. As a young person---and a woman---she did not have the standing to speak for herself in sessions of the Royal Academy. Thus did Felix Klein report her results. 
The title page of the paper he read (Figure \ref{fig:ivp}) reveals Noether's interesting approach: She combines the  notions of 
\noindent the calculus of variations (or the Euler--Lagrange equations, in more technical terms) with the theory of groups to explore what can you extract from a differential equation subjected to constraints of symmetries. Her principal results can be stated in two theorems$\,$\footnote[][12pt]{Summary of Emmy Noether's report to the German Mathematics Club, \href{http://bit.ly/2BxXXMl}{Jahresbericht der Deutschen Mathematiker-Vereinigung Mitteilungen und Nachrichten vol 27, part 2, p. 47 (1918)}.}:
\begin{enumerate}[I.]
\item If the integral $\mathscr{I}$ is invariant under a finite continuous group $\mathfrak{G}_\rho$ with $\rho$ parameters, then there are $\rho$ linearly independent combinations among the Lagrangian expressions that become divergences---and conversely, that implies the invariance of $\mathscr{I}$ under a group $\mathfrak{G}_\rho$.

\item If the integral $\mathscr{I}$ is invariant under an infinite continuous group $\mathfrak{G}^{\infty}_{\rho}$ depending on $\rho$ arbitrary functions and their derivatives up to order $\sigma$, then there are $\rho$ identities among the Lagrangian expressions and their derivatives up to order $\sigma$. Here as well the converse is valid.
\end{enumerate}

What are the implications of these propositions? Theorem~I includes all the known theorems in mechanics concerning the first integrals, including the familiar conservation laws$\,$\footnote[][0pt]{A skeletal but useful reference is \bibentry{RevModPhys.23.253}} shown in Table~\ref{tab:mechanics}. Interestingly, examples of relationships such as these were actually known in special cases before Noether's work. Theorem~II, which implies differential identities, may be described as the maximal generalization in group theory of ``general relativity.'' 
{\begingroup
 \setlength\tabcolsep{1.66mm}
\begin{margintable}[-12pt]
  \centering
  \fontfamily{ppl}\selectfont
   \caption{Symmetries and conservation laws of classical mechanics.\B}
 \begin{tabular}{lc}
    \toprule
    Symmetry & Conserved\B \\
    \midrule
    \hspace*{-8pt}\begin{tabular}{l}
    Translation in space\\
\emph{No preferred location}\end{tabular}
 & {Momentum} \\[12pt]
   \hspace*{-8pt} \begin{tabular}{l}
    Translation in time \\
\emph{No preferred time}
\end{tabular}
 & Energy \\[12pt]
   \hspace*{-8pt} \begin{tabular}{l}
    Rotation invariance\\
\emph{No preferred direction}
    \end{tabular}\qquad & \hspace*{-8pt} \begin{tabular}{c}Angular \\ Momentum
    \end{tabular} \\[12pt]
    \hspace*{-6pt}\begin{tabular}{l}
    Boost invariance\\
\emph{No preferred frame}
    \end{tabular} & C.M.\ theorem \\
    \bottomrule
  \end{tabular}
  \label{tab:mechanics}
\end{margintable}
\endgroup}
What is striking about the Theorems is their utter generality. You don't have to restrict yourself to a specific equation of motion, you don't have to stop after the first derivative, you can have as many derivatives as you want in the Lagrangian of the theory, and  you can make this generalization beyond simple transformations to more complicated ones.

To translate those Theorems into the language we physicists use with our students, Theorem I links a conservation law with every continuous symmetry transformation under which the Lagrangian is invariant in form. This is, from our perspective, a stunning development. Consider the conservation of energy. The science of mechanics developed step by step, often by inspired trial and error. Clever people made guesses about what might be a useful quantity to measure, what might be a constant of the motion. Even something as fundamental as the Law of Conservation of Energy was sort of an empirical regularity. It didn't come from anywhere, but it had been found to be a useful construct. After Noether's Theorem I, we know that energy conservation does come from somewhere that seems rather plausible: the idea that the laws of nature should be independent of time. We can derive what had appeared to be useful empirical regularities from the symmetry principles$\,$\footnote{For an example derivation, see Chapter~2 of \bibentry{Quigg:2013ufa}.}.

The distinguished group theorist Feza Gürsey, who taught physics at Yale and at the Middle East Technical University in Ankara, was rapturous about the implications. In Nathan Jacobson's introduction to Emmy Noether's collected works$\,$\cite{noether1983gesammelte}, Gürsey is quoted,  \begin{quote}Before Noether's Theorem, the principle of conservation of energy was shrouded in mystery, leading to the obscure physical systems of Mach and Ostwald.  Noether's simple and profound mathematical formulation did much to demystify physics.\end{quote}

For its part, Theorem II contains the seeds of gauge theories (``Symmetries dictate interactions'') and exhibits the kinship between general relativity (general coordinate invariance) and gauge theories. We will have more to say about the strategy of gauge theories at the end of this essay.
On the way, Noether's analysis clarified Klein and Hilbert's issue about energy conservation in General Relativity$\,$\footnote[][-56pt]{General coordinate invariance gives rise to the Bianchi identities that cause the energy conservation law to seem trivial. Energy conservation arises from the symmetry, as explained in \bibentry{Brading:2005ina}. The canonical modern treatment is \bibentry{Arnowitt:1962hi}.}.


\newthought{The person who gave us these theorems} was Amalie Emmy Noether. She was called by her middle name because both her mother and her grandmother were also named Amalie. She was born on March 23, 1882 in Erlangen, a university town a bit north of N\"urnberg. At the time of her birth, the population was about fifteen thousand. The most famous son of Erlangen was Georg Simon Ohm, the $V=IR$ lawgiver, who was not only born in Erlangen but also earned his Ph.D. there. Emmy's father, Max Noether$\,$\cite{MaxLebenWerk}, was a Professor of Mathematics at the University of Erlangen from 1875. That surely influenced her development. He did algebraic geometry, the study of curves on surfaces. Max was a scholar of some distinction, elected to the Academies of Berlin, Göttingen, Munich, Budapest, Copenhagen, Turin, Accademia dei Lincei, Institut de France, and the London Mathematical Society.

Felix Klein, whom we have already met as the person who delivered Noether's Theorems, passed through Erlangen for three years, and he had put it on the mathematical map. In his Inaugural Address (1872), he set out a research plan to study geometry from the perspective of group theory. Until that time, the basis of geometry had been to start out with rectilinear coordinate systems. Klein's innovation, with Riemann in the air, was that you shouldn't be tied to a coordinate system, or to a Euclidean space, as we would say today. Instead, it should be the symmetries of the objects that you are talking about---the group structure, and not just the $x$, $y$, and $z$ coordinates. Klein then moved on to a series of other positions, but he had left his mark with the ``Erlangen Program,''$\,$\cite[-57pt]{10.5749/j.cttttp0k.9} so the university was known to be a place that was serious about mathematics.

Max Noether was a proteg\'e and collaborator of Alfred Clebsch, and was later the intellectual executor of Clebsch's work. Clebsch had another junior collaborator named Paul Gordan, who was a colleague of Max Noether. We know the Clebsch--Gordan pair for the decomposition of combinations of angular momentum vectors$\,$\footnote{The Particle Data Group's Table of Clebsch--Gordan coefficients, \url{pdg.lbl.gov/2018/reviews/rpp2018-rev-clebsch-gordan-coefs.pdf}.}. 
Gordan was a strong presence in the Mathematics Department at Erlangen when Max was on the faculty. He is depicted as a strange fellow who would perambulate around town, smoke a cigar, stop in a beer garden, all while thinking deep thoughts. According to his colleagues, he was capable of writing down a complete paper without any pause. He is said to have written a paper in which there are twenty consecutive pages of formulas, without a single intervening word. In his obituary, written by Max and Emmy Noether, they decree that he was an \emph{Algorithmiker,} a maker of algorithms. 

What about Emmy Noether herself, how did she become a promising young mathematician$\,$\footnote{For a brief account of the early years, see \bibentry{NoetherCuz}.}? Like many young women of her milieu---aspiring middle class, with some intellectual inclinations---she attended the \emph{Städtische Höheren Töchterschule} from 1889 to 1897. Nominally, that was preparation for the life of a lady in which, if you had a profession at all, it would be teaching English and French to other young ladies. Upon completing that curriculum, she passed in 1900 the Bavarian State Exam for teachers of French and English. At the time, she could not enroll in the University of Erlangen, because women were not allowed to do that$\,$\footnote{In 1898, the Erlangen Academic Senate held that the ``admission of women would overthrow all academic order.'' See \hyperref[app:women]{the Appendix} for some examples of the integration of women into American universities.}. It was possible, however, to apply for special permission to listen to lectures. This opening came at different times in different German institutions. The Dean at Erlangen, who had permitted this great reform, was none other than Max, her father. 


While Emmy Noether was following the traditional course of ladylike study, she was taking private lessons on ``the mathematical curriculum of the humanistic \emph{Gymnasium}'' in Stuttgart and Erlangen, preparing for university studies$\,$\footnote{For a detailed account (in German), see Cordula Tollmien, 
``Das mathematische Pensum hat sie sich durch Privatunterricht angeeignet'' --- Emmy Noethers zielstrebiger Weg an die Universität,  in \emph{Mathematik und Gender} \textbf{5}, 1--12 (2016), Tagungsband zur Doppeltagung Frauen in der Mathematikgeschichte + Herbsttreffen Arbeitskreis Frauen und Mathematik (edited by Andrea Blunck, Renate Motzer, Nicola Ostwald), Franzbecker-Verlag für Didaktik  \url{http://www.cordula-tollmien.de/pdf/tollmiennoether2016.pdf}.}.
She was able to present these credentials in her October 1900 petition to attend university lectures, to establish that she was indeed prepared to benefit. 

%
%

In 1903, she passed the university qualification, but she still couldn't be admitted to the University of Erlangen. (Perhaps her father, the Dean, hadn't moved fast enough.) The University of G\"ottingen was a little bit more open-minded. She went there for a semester, during which she heard lectures by Karl Schwarzschild, Hermann Minkowski, Felix Klein, and David Hilbert. I think if you do that in your first semester of university, you're either converted---or you're history! Emmy Noether was converted---and she would make history. After one semester, Erlangen saw the error of its ways and began admitting women, precisely two out of a class of about a thousand, and so she was able to enroll at the University of Erlangen as a student of mathematics.

She wrote her dissertation in 1907 under the direction of Gordan, whom she had known throughout her childhood. Emmy Noether received her D. Phil. \emph{summa cum laude} for ``Über die Bildung des Formensystems der ternären biquadratischen Form'' (On the construction of the system of forms of a ternary quartic form). The work involved the meticulous computation of some 331 invariants of quartic forms---a very Gordanian undertaking. She later described her thesis topic as \emph{Mist} (dung), hardly an expression of pride, because she aspired to invent, not merely to compute. It appears that Noether was the second woman Ph.D. in mathematics in Europe, following Sofia Kovalevskaya who received her degree in Göttingen in 1874, with Karl Weierstra\ss, rose to full professor at Stockholm 1889, and died aged forty-one in 1891.

\emph{Dr.} Emmy Noether remained in her home town as an unpaid member of the Erlangen Mathematical Institute from 1908 to 1915. She got a lot of experience in teaching and research. When her father's health began to fail, she took over his classes. She was conducting herself as a faculty member, though without either compensation or status. She became a member of the German Mathematical Society (\emph{Deutsche Mathematiker-Vereinigung}) in 1909, and in the same year became the first woman to lecture at the Society's annual meeting. The department added new faculty members, and she came under the influence of Ernst Fischer$\,$\cite{ernstfischer}---Paul Gordan's successor---who gave her an entry into the world of abstract mathematics, rather than mere computation. It was for abstract mathematics that she turned out to have an enormous talent. 

\newthought{In 1915, she was invited to Göttingen} by Klein and Hilbert. Göttingen was the Mount Olympus$\,$\footnote{Benno Artmann,
``Hochburg der Mathematik,'' in \emph{Georgia Augusta} (2008) \url{http://bit.ly/2GQmQZL}, pp.~14--23.} of mathematics, at least in Germany. It was  where Carl Friedrich Gau\ss\ had held court. If you look at the list of heroes on their history page, you will find many familiar names: Carathéodory, Clebsch, Richard Courant, Dirichlet, Herglotz, Kästner, Minkowski, Carl Runge, and  Hermann Weyl, among others. It was a great place to be a young person in mathematics.

Göttingen's proud tradition in mathematics includes an unequalled trove of information about the early history of modern (eighteenth and nineteenth century) mathematics. A locked \emph{Giftschrank} (poison cabinet) in the math library holds treasures including notes$\,$\footnote[][-15pt]{Felix Klein, \emph{Seminar-Protokolle,} \url{http://www.claymath.org/publications/klein-protokolle}. For a brief tour, see Eugene Chislenko and Yuri Tschinkel, ``The Felix Klein Protocols,'' Notices Amer. Math. Soc. \textbf{54}, 961--970 (2007), \url{http://www.ams.org/notices/200708/tx070800960p.pdf}.} of forty years of seminar lectures by Felix Klein, his colleagues and students, and distinguished visitors---eight thousand pages in twenty-nine volumes!  

Hilbert took a very keen interest in Emmy Noether, and worked to advance her career. 
The Mathematics and Science Department of the Philosophical Faculty put her forward in 1915 for the \emph{Habilitation} lecture to become Privatdozent in Göttingen, with unanimous---if somewhat old-school---support. One endorsement came from Göttingen mathematician Edmund Landau$\,$\cite{Schappa}:
\begin{quote}
I have had up to now uniformly unsatisfactory experiences with female students and I hold that the female brain is unsuited to mathematical production. Miss Noether seems to be a rare exception.
\end{quote}
However, in a special vote against the \emph{Habilitation} of Emmy Noether, 19 November 1915, the Historical-Philological Department blocked the move, out of ``concern that seeing a female organism might be distracting to the students.''$\,$\footnote{For the full German text, see Cordula Tollmien, ``Weibliches Genie: Frau und Mathematiker: Emmy Noether,'' in \emph{Georgia Augusta} (2008) \url{http://bit.ly/2GQmQZL}, pp.~38--44.}


The \emph{Habilitation} was not formally refused by the university; the administration simply never took action. 
Accordingly, the \emph{Habilitation} was not granted. But with Hilbert as her patron, Emmy Noether was  permitted to lecture under his name. Courses were announced under Hilbert's authority, with the assistance of Fr\"aulein Noether. He might appear at the first class and the last, and everything else was in her care. She received no official compensation for her service, but there are hints that some arrangement might have been made.

The G\"ottingen mathematicians pressed her case again in 1917, this time with new urgency: the fear that she would be hired away to Frankfurt if G\"ottingen did not go ahead. They applied to the Ministry to make an exception, to save this talent that was indispensable to G\"ottingen. The reply from the Ministry of Education$\,$\footnote{Letter from the Ministry of Education, the Edelstein Collection, the National Library of Israel, \url{http://bit.ly/2BFZHDs}. English translation at \url{https://blog.nli.org.il/en/noether/}.} exhibits unimpeachable bureaucratic logic.\vspace*{-6pt}
\begin{quotation}
\phantom{M}\hfill Berlin, July 20, 1917\phantom{MMM}\break
With regard to accepting women to teaching positions, the regulations of Frankfurt University are identical to those of all the universities: women are not allowed to be appointed to positions of external lecturers. It is completely impossible to make an exception to the rule in one university. Therefore, your concern that Miss Noether will leave, move to Frankfurt and receive a position there is completely unfounded: she will not be given the right to teach there, just as she will not receive such a thing in Göttingen or in any other university. The Minister of Education has expressed this time and time again and emphasized that he supports his predecessor's instructions, and therefore women will not be permitted to receive teaching positions in universities.\\[6pt]
\noindent Therefore, there is no concern that you will lose Miss Noether as an external lecturer in Frankfurt University.
\end{quotation} 

The foundation of the Weimar Republic, following Germany's defeat in the War of 1914--1918, brought liberalization and many reforms: Women were no longer explicitly forbidden to teach in universities. In 1919, Emmy Noether was granted her \emph{Habilitation} on the basis of her paper on ``Invariant Variational Problems.'' She was now an adjunct faculty member of a sort, again with no documented pay for her services. 


\newthought{Might symmetries beget interactions?} 
One of Emmy Noether's colleagues, a frequent visitor to G\"ottingen who eventually took a position there, was Hermann Weyl, one of the pioneers of the application of symmetries to modern physics. Weyl had an interesting idea---also in 1918, the year of Noether's theorems. He set out to make a unified theory of all the fundamental interactions then known: electromagnetism and gravitation$\,$\cite{Weyl:1918ib}. He had the notion that he might derive this theory from a symmetry principle, by building a theory that was invariant under scale transformations. Imagine that measuring sticks change scale as a function of position, and require that the theory be invariant under those scale changes. The construction failed as a physical theory$\,$\footnote{See \S3.1 of \bibentry{Kastrup:2008jn}.}. It did not lead to Maxwell's equations and, on the gravitation side, Einstein himself objected that the way clocks tick would depend on the path traversed from one point to another. So, Weyl's is a wrong idea, but as with many ``wrong'' ideas in physics, there is something very clever about it: the idea that interactions might be derived from symmetries$\,$\footnote{See \bibentry{Wu:2006dt} for an illuminating historical treatment.}.

No one at the time noticed the connection between Weyl's notion and Noether's second theorem, which we now understand shows that such a construction is always possible. Part of the reason is that certain other pieces were missing. After the invention of quantum mechanics, and the eventful decade that followed, with prodding from Einstein and Fock and others, Weyl came to the realization that he could indeed derive electrodynamics from a symmetry principle by imposing a certain symmetry on the wave function---an essential new feature of quantum mechanics. We prove in our beginning courses that the absolute phase of a quantum-mechanical wave function is a matter of convention, with no observable consequences. If you go further, and impose freedom to choose the phase convention independently at every point, in the style of the second theorem, you can derive electrodynamics from the Schr\"odinger equation. 

In 1931, in the paper in which he invented quantum electrodynamics and the monopole$\,$\cite{Dirac:1931kp}, Dirac talks somewhat mystically about something he called the \emph{nonintegrable phase.} In classical electrodynamics, we know that potentials contain too much information, and it was long believed that the electric and magnetic fields contain all the information needed. That turns out to be incorrect: in quantum mechanics, the fields contain too little information. There is an intermediate, path-dependent phase factor that is both nonlocal and topological, that contains just the right amount of information, as explained in 1959 by Aharonov and Bohm$\,$\cite[-41pt]{Aharonov:1959fk}. 

Later in life (1955), trying to explain how he knew he was on the right track, Weyl wrote$\,$\footnote{Quoted in Freeman J. Dyson, \emph{Birds and Frogs: Selected Papers of Freeman Dyson, 1990--2014,} World Scientific, Singapore, 2015, p.~47.},
\begin{quote}
The strongest argument for my theory seemed to be this: the gauge invariance corresponds to the principle of conservation of electric charge as the coordinate invariance corresponds to the conservation law of energy and momentum.
\end{quote}
I interpret this to mean that somehow, either explicitly or vaguely in his mind was an understanding of Noether's theorem and the connection between symmetries and conservation laws.

\newthought{A central feature of electrodynamics} is that electric charge is conserved. The best current limit on charge conservation comes from the Borexino experiment$\,$\cite{Agostini:2015oze}, an exquisitely radiopure liquid scintillation detector  located deep underground at the Gran Sasso Laboratory. They derive a new limit on the stability of the electron for decay into a neutrino and a single monoenergetic photon. This new bound, $\tau \ge 6.6 \times 10^{28}\hbox{ yr}$ at 90\% C.L. improves the previous limit by two orders of magnitude.

Where does charge conservation come from? Why is charge conserved? You might say that it is implied by Maxwell's equations. But if you look back at how Maxwell formulated his equations, out of the observations of Faraday, he tuned them so that charge would turn out to be conserved under all circumstances. That's where the displacement current comes from, as an addition to Ampère's law in nonstatic cases. That is to say, the Maxwell equations were built to explain the observation that electric charge is conserved. So to say that charge conservation follows from Maxwell's equations is not a deep explanation, although it serves us pretty well in most circumstances.

We can use the global phase invariance of Theorem~I to imply the existence of a conserved charge that we identify as the electric charge. This is an important step toward a derivation, but we might just as well identify that conserved charge as baryon number, for example. To my mind, only when we apply the local phase invariance of Theorem~II and show that the theory that results is indeed electromagnetism, can we be certain that the charge we have defined is the electric charge. There is still a coupling constant here, and you must still identify that coupling with the electric charge, but you have derived the whole form of Maxwell's equations, so it's not much of a leap.

From this notion that you should be able to choose the phase convention independently at every seat in the auditorium, we can derive (the Lagrangian and equations of motion of) quantum electrodynamics, and therefore charge conservation$\,$\footnote[][-70pt]{For further discussion, see \bibentry{BRADING20023}.}.  In analogy with the kinematic conservation laws, this is a step in pushing back the origin of the conservation laws by showing that they can be derived from a symmetry principle. They are not merely empirical regularities. Now, the exactness of the symmetry principle can still be challenged, and you can make an unfortunate choice of a symmetry principle, but Noether's theorems give a deeper understanding of why the conservation laws should hold.

\newthought{Noether's ``Invariant Variational Problems'' made waves} in general relativity circles, but wasn't otherwise an instant sensation. It was what our friends at \url{inspirehep.net} call a ``Sleeping Beauty.'' Werner Heisenberg was a great proponent of symmetry in fundamental physics. (He was, after all, to be the inventor of isospin.) Late in life, holding court with his disciples about the Meaning of Everything, he made this ringing pronouncement$\,$\cite{WernerH}:
\begin{quote}
``In the beginning was the symmetry,'' that is certainly more correct than the Democritean thesis, ``in the beginning was the particle.'' The elementary particles embody the symmetries, they are their simplest representations, but they are above all a consequence of the symmetries.
\end{quote} 
It is not certain, but there is evidence from other interviews that he never read Noether's paper: ``[I]t did not penetrate into quantum theory, so I didn't realize the importance to that paper.''$\,$\footnote{See pp. 85--86 of \emph{The Noether Theorems,} Ref.~2.} I suspect that Heisenberg and his cohort had plenty to do---inventing and applying quantum mechanics---and that once they had heard of the obvious consequences of Noether's theorem---the conservation laws of mechanics---they surmised that they already knew that, and had no need to pay attention. The other important point was that internal symmetries had yet to be invented. (From our point of view, you apply the theorems to internal symmetries to make gauge theories.) Internal symmetries such as isospin did not exist, would not be invented until after the discovery of the neutron in 1932.

Emmy Noether was thus not instantly revered in the community of physics. Some have speculated that the ferment in both quantum physics and modern algebra in G\"ottingen was so consuming for the physicists and mathematicians, respectively, that they did not notice the mutual relevance of their new developments. Since the 1960s, ``Invariant Variational Problems'' and Emmy Noether have been having a moment, so to say, among physicists and others$\,$\footnote[][-24pt]{Crowned by a Google doodle: \url{https://www.google.com/doodles/emmy-noethers-133rd-birthday}.}.

You may have heard of the famous suggestion by Niels Bohr$\,$\cite[]{JR9320000349} that the continuous $\beta$-decay spectrum might be explained by the hypothesis that for microscopic phenomena energy conservation could be a statistical phenomenon and not a rigid law:\newpage
\begin{quote}
At the present stage of atomic theory we have no argument, either empirical or theoretical, for upholding the energy principle in $\beta$-ray disintegrations, and are even led to complications and difficulties in trying to do so.
\end{quote}
 That was not the first time that he had explored deviations from strict energy conservation. A 1924 paper by Bohr, Kramers, and Slater$\,$\cite{BKS1924} raised the possibility that for radiative processes and on a small scale energy conservation might be enforced in some statistical sense. While many physicists objected$\,$\footnote{For a commentary, see \bibentry{Kragh2009}.}, no one seems to have invoked Noether's insight to say, ``There's a theorem. You can't do this," or at least "You'd be paying a big price." The conjecture was buried within a year by precise measurements of the final-state momenta in Compton scattering. 

\newthought{In the hotbed of G\"ottingen,} Emmy Noether's approach to mathematics changed. She stopped doing computation and became interested in Modern Algebra. The famous booklet$\,$\cite{Galois1846} of \'Evariste Galois, applying group theory to the solution of algebraic equations, was an inspiration. With Hilbert's support, she was appointed Au\ss erordentlicher Professor (a real adjunct position, but again without pay from the university) in 1922. Hilbert was able to provide a small stipend, and she had some family money. She had robust connections with the Soviet mathematicians and spent the year 1928--1929 at Moscow State University. She did spend a little time in Frankfurt in 1930, although she was not hired away.

Recognition began to come her way. In 1932, she received the Alfred Ackermann-Teubner Award together with her collaborator Emil Artin$\,$\cite{artin1998galois}, another pioneer of algebraic equations. In that same year, Noether was the first woman invited to give a Plenary lecture at International Congress of Mathematicians, held in Zürich. She was a very devoted editor of the \emph{Mathematische Annalen.} 

Emmy Noether was, by many testimonies, the hub of activity in G\"ottingen. She had a devoted following of students and young collaborators, mostly men, called \emph{die Noetherknaben} (the Noether boys). They were said to roam around G\"ottingen in an unruly mass, debating mathematics. It provoked a minor scandal when they prowled the town, not wearing coats and ties, although the only photographs I have come across do show them in coats and ties. Many of them grew up to be distinguished and quite well-known mathematicians. Hermann Weyl later confessed that the mathematical men of G\"ottingen referred to Emmy as \emph{Der Noether,} in the masculine---in the retelling a term of respect because she was as strong as any man in mathematics. Her pioneering studies of rings and ideals earned her a less ambiguous title: The Mother of Modern Algebra.

\newthought{In 1933, an edict appeared} in the name of the Culture Minister, the notorious Bernhard Rust, announcing that anyone of Jewish background had to be put on leave from the university$\,$\footnote[][-12pt]{For an account from the perspective of six decades, see \bibentry{SMLnazis}.}. According to an account in the \emph{G\"ottinger Tageblatt} of April 26$\,$\footnote{\url{http://www.tollmien.com/noethertelegrammapril1933.html.}}, Emmy Noether was among the first six faculty members rusticated from the University. The others in mathematics and physics were Felix Bernstein (a founder of biostatistics), Max Born (who would receive the 1954 Nobel Prize for his statistical interpretation of quantum mechanics), and Richard Courant. Courant was by that time running the institute, succeeding Hilbert, who had passed the obligatory retirement age of sixty-eight. The two collaborated on the famous Courant--Hilbert volumes on mathematical physics$\,$\cite{9780470179529} in which, by the way, the Noether theorems are discussed (Ch.~IV, \S12.8). 
Among the Nazi informants was Werner Weber, one of Emmy Noether's doctoral students. 

The order to take a leave of absence was an ominous development, and the implications soon became inescapable. On May 10, 1933, German students incinerated tens of thousands of ``un-German'' books in Berlin's \emph{Opernplatz} and in G\"ottingen and other university towns. Leaders of twenty-one American universities and colleges moved quickly to establish an Emergency Committee in Aid of Displaced German Scholars. The operating officer of the Emergency Committee was one Edward R. Murrow---before he was a legendary newsman$\,$\cite[-60pt]{MurrowIIE}. 

In September came a communication in the form of a telegram$\,$\footnote{Several interesting documents from the Edelstein Collection in the National Library of Israel appear in \bibentry{ENNatLibIsrael}.} from the Prussian Ministry of Science, Art, and Education in Berlin saying that on the basis of \S3 of the Law for the Restoration of the Professional Civil Service of April 1933, Emmy Noether's teaching permit was nullified. The university was instructed that her wages, such as they were, were to cease by the end of the month. 

Sympathetic colleagues, Hilbert among them, had to scramble to find landing places for Emmy Noether and many others---by the end of 1933 eighteen mathematicians left or were driven out from the faculty at the Mathematical Institute in Göttingen alone.
 Born went off to the University of Cambridge and Bangalore before settling in Edinburgh as Tait Professor of Natural Philosophy$\,$\cite[-24pt]{born1978my}. Richard Courant found his way via Cambridge to New York$\,$\footnote{An extensive discussion of the drama of 1933 appears in chapters 15 and 16 of \bibentry{reid1976courant}.}, where he founded what is now the Courant Institute for Mathematical Sciences at New York University. 
 
 Emmy Noether was offered a two-year Visiting Professorship at Bryn Mawr College in Pennsylvania. 
 Bryn Mawr was founded in 1885, among the earliest women's colleges established in the United States to open higher education to women. It offered rigorous intellectual training, including postgraduate study, and the opportunity to engage in original research in the tradition of European universities.
 In a brief announcement of her appointment, The New York \emph{Times}, with a delicacy occasionally in evidence today, reported that ``She was asked, with other members of the G\"ottingen faculty, to resign last spring, under the Nazi regime.''$\,$\cite[]{nyt1933}

The President of Bryn Mawr and  Emmy Noether's supporters back in Germany recognized that although she had a mastery of English---and a certificate to show it---perhaps she wasn't ideally suited to undergraduate instruction. Bryn Mawr already had a small graduate program in mathematics for which she was an ideal fit.  To make the most of the celebrated mathematician's presence, the college enlarged the circle of women in mathematics by creating Emmy Noether scholarships and fellowships$\,$\footnote{Four of her Bryn Mawr students and Emmy Noether Fellows have contributed admiring recollections: \bibentry{NoetherGirls}. For additional information about Noether's association with Bryn Mawr, see \bibentry{Shen2019}.}. It was arranged, moreover, that she could take weekly trips to the Institute for Advanced Study where she gave seminars and lecture courses. The Institute was already becoming one of the great centers of mathematical research. That affiliation gave her contact with other illustrious immigrants, among them colleagues she had known in Germany, such as Weyl. Einstein had noticed her work, but it is not clear that they ever had any real contact. 

Whether or not the Bryn Mawr women following ``Miss Noether'' on brisk walks were as disorderly as the \emph{Noetherknaben} in G\"ottingen, they appeared to be just as engaged and effervescent. Emmy Noether herself was looking forward, full of curiosity about American ways, stimulated by her students and her interactions at Princeton, and generally full of life. During the spring vacation of 1935, she had abdominal surgery that was expected to be routine. She seemed to be recuperating well, but suffered complications and  died within a few days.  

Einstein sent a letter of eulogy to the \emph{Times}$\,$\cite{AEonEN}. She was, he wrote, 
\begin{quote}\ldots the most significant creative mathematical genius thus far produced since the higher education of women was begun. In the realm of algebra, in which the most
gifted mathematicians have been busy for centuries, she discovered methods which have proved of enormous importance in the development of the present-day younger generation of mathematicians. 
\end{quote}
At Bryn Mawr, her ashes were interred under the walk in the cloisters, beneath  a modest marker bearing the imprint E $\cdot$ N 1882--1935.

Speaking at her memorial service at Bryn Mawr, Hermann Weyl gave a very long, and very detailed, hymn of praise$\,$\footnote{Reprinted in \hyperref[auguste]{Auguste Dick's \emph{Emmy Noether, 1882-1935}, pp. 112--152}.}.
\begin{quote}
I have a vivid recollection of her when I was in Göttingen as visiting professor in the winter semester of 1926-1927, and lectured on representations of continuous groups. She was in the audience; for just at that time the hypercomplex number systems and their representations had caught her interest and I remember many discussions when I walked home after the lectures, with her and von Neumann, who was in Göttingen as a Rockefeller Fellow, through the cold, dirty, rain-wet streets of G\"ottingen. When I was called permanently to G\"ottingen in 1930, I earnestly tried to obtain from the Ministerium a better position for her, because I was ashamed to occupy such a preferred position beside her whom I knew to be my superior as a mathematician in many respects.
\end{quote}

Pavel Alexandrov in Moscow was one of Emmy Noether's closest friends. Part of her fascination with the Russian school of mathematics was channeled through him. He writes very respectfully and lovingly about what a wonderful person she was, how generous to her students$\,$\footnote{Reprinted in \hyperref[auguste]{Auguste Dick's \emph{Emmy Noether, 1882-1935}, pp. 153--179}.}. Apparently she would throw out ideas, elaborate programs, let her students carry them out, see that her students would write them up and get credit for them.
\begin{quote}
With the death of Emmy Noether I lost the acquaintance of one of the most captivating human beings I have ever known. Her extraordinary kindness of heart, alien to any affectation or insincerity; her cheerfulness and simplicity; her ability to ignore everything that was unimportant in life---created around her an atmosphere of warmth, peace and good will which could never be forgotten by those who associated with her.  \ldots Though mild and forgiving, her nature was also passionate, temperamental, and strong-willed; she always stated her opinions forthrightly, and did not fear objections. It was moving to see her love for her students, who comprised the basic milieu in which she lived and replaced the family she did not have. Her concern for her students' needs, both scientific and worldly, her sensitivity and responsiveness, were rare qualities. Her great sense of humor, which made both her public appearances and informal association with her especially pleasant, enabled her to deal lightly and without ill will with all of the injustices and absurdities which befell her in her academic career. Instead of taking offense in these situations, she laughed. 
\end{quote} 

Bartel van der Waerden, who synthesized$\,$\cite{waerden1991algebra} the exciting lectures of Emil Artin and Emmy Noether in which they were creating the axiomatic approach  to modern algebra, wrote$\,$\footnote{Reprinted in \hyperref[auguste]{Auguste Dick's \emph{Emmy Noether, 1882-1935}, pp. 100--111}.}  
\begin{quote}
This entirely non-visual and noncalculative mind of hers was probably one of the main reasons why her lectures were difficult to follow. She was without didactic talent, and the touching efforts she made to clarify her statements, even before she had finished pronouncing them, by rapidly adding explanations, tended to produce the opposite effect. And yet, how profound the impact of her lecturing was. Her small, loyal audience, usually consisting of a few advanced students and often of an equal number of professors and guests, had to strain enormously in order to follow her. Yet those who succeeded gained far more than they would have from the most polished lecture. She almost never presented completed theories; usually they were in the process of being developed. Each of her lectures was a program. And no one was happier than she herself when this program was carried out by her students. Entirely free of egotism and vanity she never asked anything for herself but first of all fostered the work of her students. She always wrote the introductions to our papers \ldots
\end{quote}
\noindent Van der Waerden has written elsewhere that when they went walking in G\"ottingen, as she did with her students at Bryn Mawr, Emmy Noether would talk so rapidly and with such excitement as to be utterly incomprehensible. It came to him that if he led her on several laps around the city, she became, by the third lap, slightly short of breath and spoke slowly enough that he could understand her.

\newthought{Postscript.} Two decades would pass after Emmy Noether's death before physicists began to exploit the full power of Theorem~II. The notion that internal symmetries could generate interactions was put into practice by Yang and Mills$\,$\cite{Yang:1954ek}, who sought to derive a theory of the strong interactions among nucleons from isospin symmetry. They asked whether it should not be possible to choose the isospin convention independently at every point in space time, much as we set the phase convention of the quantum mechanical wave function locally to derive quantum electrodynamics. The mathematical construction goes through: the symmetry implies a conserved isospin current, and massless vector fields that interact among themselves mediate the forces between nucleons. This doesn't correspond to the real world. As with many ideas in physics, the first time it is applied it doesn't work but the idea remains. And we have now found how to apply the idea successfully---in the theory of quantum chromodynamics for the strong interactions among quarks and gluons, and in the electroweak theory, where the gauge symmetry must be hidden. 

\vspace*{-6pt}
\section{Appendix: The Arrival of Women in American Universities\label{app:women}}
Lest we imagine that the German universities were singularly backward in granting women admission as students and appointments as faculty members, let us briefly review the American experience.

The universities established during the colonial period in what became the United States were closed to women students for many years. Generally speaking, higher education was not available to women in this country until the nineteenth century, when women's colleges were created to provide rigorous academic training on a par to what was reserved to men. Graduate degrees for women remained a rarity well into the twentieth century$\,$\cite[-48pt]{10.2307/367747,10.2307/40222081}. The special role of women's colleges in providing training and academic employment has persisted long after women were accepted into graduate programs in the major universities$\,$\cite{10.1088/978-1-6817-4094-2ch5}.

It is instructive to consider the record of some leading American institutions.
\vspace*{-6pt}
\begin{quote}
Princeton University first admitted women as undergraduates in 1969, eight years after the first full-time woman graduate student. The first female physics graduate student received her Ph.D.  in 1971. The first woman was appointed as a tenured professor in 1968, and the first woman was tenured in physics in 1998.

Yale University admitted women as graduate students beginning in 1892, and the first science Ph.D.s were awarded to women in astronomy and chemistry in 1894. The first Ph.D. awarded to a woman in physics was in 1932. Women have served on the Yale faculty since 1920, but none was tenured until the 1950s. Undergraduate women were admitted in 1969. The first woman was tenured in physics in 2001, the first in mathematics in 2003.

The University of California at Berkeley admitted women on an equal basis with men from 1870, two years after its founding. A woman first earned a physics Ph.D. in 1926. The first woman on the physics faculty was appointed, with tenure, in 1981. 

\end{quote}

\vspace*{-12pt}

\section{Additional Sources\label{sources}}
\begin{quotation}
\bibentry{dick1981emmy}.\label{auguste}\\[6pt]

\noindent\bibentry{noether1981emmy}.\\[6pt]

\noindent\bibentry{noether1983emmy}.\\[6pt]

\noindent\bibentry{Kastrup}.\\[6pt]

\noindent \bibentry{lederman2008symmetry}.
\\[6pt]

\noindent\bibentry{IASEmmy}; \bibentry{ENParadise}.\\[6pt]
\noindent\bibentry{Olver}.\\[6pt]
\noindent\bibentry{10.2307/27962871}.\\[6pt]
\noindent\bibentry{WomenMath}.\\[6pt]
\noindent Slides illustrating the colloquium on which this article is based are available at \bibentry{quigg_chris_2018_1346275}.
\end{quotation}
\nobibliography{emmyV2}
\bibliographystyle{plainnat-eprints}

\end{document}